\documentstyle[aps,prb,multicol,epsf]{revtex}

\begin{document}

\title{Analysis of negative magnetoresistance. \\
Statistics of closed paths. II. Experiment}

\author{G.\ M.\ Minkov,\cite{cont} S.\ A.\ Negashev, O.\ E.\ Rut, and A.\ V.\ Germanenko}
\address{Institute of Physics and Applied Mathematics, Ural State University,
620083 Ekaterinburg, Russia}

\author{O.\ I.\ Khrykin, V.\ I.\ Shashkin and V. M. Danil'tsev}
\address{Institute fo Physics of Microstructures of RSA, 603600 N.
Novgorod, Russia}
\date{\today}
\maketitle\widetext
\begin{abstract}
It is shown that a new kind of information can be extracted from the Fourier
transform of negative magnetoresistance in 2D semiconductor structures. The
procedure proposed provides the information on the area distribution function
of closed paths and on the area dependence of the  average length of closed
paths. Based on this line of attack the method of analysis of the negative
magnetoresistance is suggested. The method  has been used to process the
experimental data on negative magnetoresistance in 2D structures with
different relations between the momentum and phase relaxation times. It is
demonstrated this fact leads to distinction in the area dependence of the
average length of closed paths.

\end{abstract}
\draft
\begin{multicols}{2}
\narrowtext
\section{Introduction}
\label{secIn} The phenomenon of anomalous magnetoresistance at low temperature
in ``dirty" metals and doped semiconductors was explained by the theory of
quantum corrections to the conductivity.\cite{1,2,3} The interference
correction to the conductivity gives the main contribution to the negative
magnetoresistance in 2D structures at low temperature and low magnetic field.
A unique analytical expression for the magnetic field dependence of negative
magnetoresistance has been found in Ref.\ \onlinecite{1}
\begin{eqnarray}
\Delta \sigma (B)&=& \sigma (B)-\sigma (0) \nonumber\\
 &=&a\ G_{0} \left(\Psi
\left(0.5+\frac{B_{tr} \tau}{B\ \tau _{\varphi }
}\right)-\ln\left(\frac{B_{tr} \tau}{B\ \tau_{\varphi } }\right)\right),
\label{eq3}
\end{eqnarray}
where $G_{0} =e^{2}/(2\pi^{2}\hbar)$,  $B_{tr} =\hbar c/(2el^{2}) $, $l$ is
the mean free path, $\Psi(x)$ is a digamma function, $\tau$ and $\tau
_{\varphi}$ stand for the elastic scattering and phase breaking time,
respectively. The parameter $a$ is equal to unity for the non-interacting
case. This expression was obtained in the diffusion approximation for the
isotropic scattering by randomly distributed scatterers with a short-range
potential. Nevertheless it is universally used in the analysis of experimental
data to extract the phase breaking time and its temperature dependence through
the fitting of experimental curves. It should be pointed out that $\tau
_{\varphi }$ determined in this way is a fitting parameter rather than the
phase breaking time because some deviation of experimental curves from Eq.\
(\ref{eq3}) takes place in almost without exception. This deviation may result
from some correlations in distribution of scatterers, scattering anisotropy,
or long-range potential fluctuations in real 2D systems. This may be a
possible reason of the saturation of $\tau _{\varphi }$ with decreasing
temperature as it is seen in some experiments.\cite{102}

In the previous paper we have developed a new approach to the analysis of
anomalous magnetoresistance which makes it possible to obtain the information
about the statistics of closed paths from the magnetic field dependence of
magnetoresistance. This approach provides the basis for a new method of
analysis of negative magnetoresistance due to weak localization suppression.
In the present paper we demonstrate the potentials of this method as it is
applied for interpretation of concrete experimental results obtained for
GaAs/InGaAs/GaAs quantum wells.

\section{Basis of method}
\label{sec2} The essence of the method is clear from Eq.\ (8) of the previous
paper. \cite{prev} One can see that the Fourier transform of negative
magnetoresistance is given by
\begin{eqnarray}
 \Phi (S)&=&\frac{1}{\Phi_0}\int_{-\infty}^{\infty}dB\ \delta\sigma(B)
 \cos\left(\frac{2\pi B S}{\Phi_0}\right)= \nonumber \\
&=& 2\pi l^{2} G_{0} W(S)\exp \left( -\frac{ \overline{L}
(S,l_{\varphi } )}{l_{\varphi } } \right), \label{eq6}
\end{eqnarray}
$\Phi _{0} =2\pi c\hbar/e$  is the elementary flux quantum, $l_{\varphi}=v_F
\tau_\varphi$, $W(S)$ and $\overline{L} (S,l_{\varphi } )$ are the area
distribution function of closed paths and  the area dependence of the average
length of closed paths respectively, introduced in Section II of Ref.\
\onlinecite{prev}.

Thus, it is clearly seen from Eq.\ (\ref{eq6}) that $\Phi(S)$ contains the
information on the area distribution function of closed paths $W(S)$, and on
the function $\overline{L} (S,l_{\varphi } )$ . If $l_{\varphi }$ tends to
infinity when $T\to 0$, the extrapolation of $\Phi(S,T)$ to $T=0$ gives the
value of $ 2\pi l^{2} G_{0}W(S)$.

To determine the area dependence of $\overline{L}$ we assume that for actual
areas $\overline{L}$ is a power function of area,
$\overline{L}(S,l_{\varphi})=S^\beta f(l_{\varphi })$. The numerical
calculations of the function $\overline{L} (S,l_{\varphi })$ (see Fig.\ 4 in
Ref.\ \onlinecite{prev}) have shown that this assumption is valid in a wide
range of $S$, $l_{\varphi }$. In the diffusion approximation (i.e. for
$\tau/\tau_\varphi\ll 1$) the value of $\beta$ is about $0.67$. It is clear
that in this approximation the value of $\beta$ is independent of scattering
anisotropy. Beyond the diffusion approximation the value  of $\beta$ is lower
and depends on $\tau/\tau_\varphi$ ratio.

To extract the value of $\beta$ from experimental data one can measure $\delta
\sigma (B)$  at two temperatures, i.e. at different $l_{\varphi }$, then find
the function
\begin{equation}
A(S)\equiv \ln \left[ \frac{\Phi (S,T_{1} )}{\Phi (S,T_{2} )}
\right] =S^{\beta} (f(l_{\varphi }^{T_1} )-f(l_{\varphi }^{T_2}))
\label{eq7}
\end{equation}
and finally determine $\beta$ from $A(S)$ {\bf curve}.

\begin{figure}
 \epsfclipon
 \epsfxsize=\linewidth
 \epsfbox{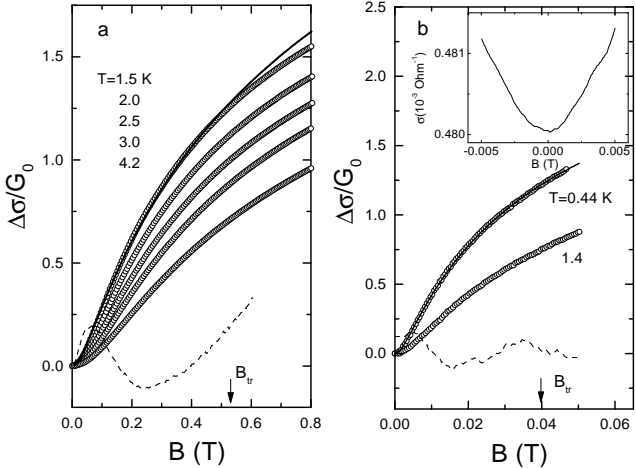}
\caption{(a) Magnetic field dependencies of $\Delta\sigma(B)/G_0$ for
different temperatures for structures I\ (a) and II\ (b). Solid lines are the
results of fitting by expression (\protect\ref{eq3}). Dashed lines are the
differences between experimental and theoretical curves multiplied by $10$ (a)
and $20$ (b). Insert in (b) shows the low field magnetoresistance at $T=1.6$ K
for structure II.} \label{fig1}
\end{figure}

\section{Experiment}
\label{sec1}

We have measured the conductivity in heterostructures
n-GaAs/In$_{0.07}$Ga$_{0.93}$ As/n-GaAs of two types. The heterostructures
with 200 \AA \ In$_{0.07}$Ga$_{0.93}$As quantum well, $\delta$-doped by Si in
the centre, relate to the first type. The heterostructures with 50 \AA \
In$_{0.07}$Ga$_{0.93}$As well and doped barriers relate to the second type.
The $\delta$-doped by Si layers are arranged in them on both sides of the well
at the distance 100 \AA. In this paper we present the experimental results for
two structures of different types which are refereed as structure I and
structure II, respectively. The measurements carried out in wide ranges of
magnetic fields (up to 6 T) and temperatures (0.4-40 K) show that in
structures investigated only one size-quantized subband is occupied. The main
contribution to the conductivity comes from the electrons in the quantum well
of In$_{0.07}$Ga$_{0.93}$As. The electron density and mobility for structure I
are $n=1.2\times 10^{12}$ cm$^{-2}$ and $\mu=1.4\times 10^3$ cm${^2}$/(V sec),
respectively. For structure II they are the following $n=2.5\times 10^{11}$
cm$^{-2}$ and $\mu=1.1\times 10^4$ cm$^2$/(V sec).

The magnetic field dependencies of the conductivity for both structures for
low magnetic fields and different temperatures are shown in Fig.\ \ref{fig1}.
The negative magnetoresistance is observed in the whole range of magnetic
fields (up to 6 T). The main contribution to the negative magnetoresistance in
the range $B<0.5$ T for structure I and  $B<0.2$ T for structure II comes from
the weak localization effect, while in the range $B>1$ T for structure I and
$B>0.4$ T for structure II it results from the correction to the conductivity
due to electron-electron interaction. Notice that the positive
magnetoresistance due to weak antilocalization effect was observed in
analogous structures for low magnetic fields in Refs.
\onlinecite{Knap,Polyanskaya}. This effect is observed when the spin
relaxation time $\tau_s$ is less than the phase relaxation time. The positive
magnetoresistance is absent in both our structures (for example, see the inset
in Fig.\ \ref{fig1}b, where magnetoresistance of structure II for very low
magnetic fields is shown). It should be mentioned that the conductivity of the
structures studied in Ref. \onlinecite{Knap} was order of magnitude larger
than that in our case therefore the phase relaxation time was longer, too. The
phase relaxation time in our case lies in the range $(0.2-1.5)\times 10^{-11}$
sec (see below). Comparing this value with $\tau_s=(3-4)\times10^{-11}$ sec
determined in Ref. \onlinecite{Knap} we have $\tau _{s}>\tau_{\varphi}$ for
our structures. Another reason for the absence of positive magnetoresistance
in our case is the fact that, in contrast to structures investigated in Ref.
\onlinecite{Knap,Polyanskaya} our structures are symmetric in the growth
direction.

Usually the expression (\ref{eq3}) is used to analyze the negative
magnetoresistance, taking $a$ and $\tau _{\varphi }$  as fitting parameters.
The solid curves in Fig.\ \ref{fig1} have been obtained in this way and, at
first glance, they are in good agreement with the experimental data in the
range of magnetic field $B<B_{tr}$. For structure I this procedure gives
$a=1$, $\tau _{\varphi }=1.25\times 10^{-11}$ sec for $T=1.5$ K. However, the
more detailed analysis reveals the difference between the theory and the
experimental data (dashed curve in Fig.\ \ref{fig1}a). As a consequence, the
parameters $a$ and $\tau_{\varphi}$ vary in the range of $0.81-1.15$ and
$(1.05 - 1.6)\times 10^{-11}$ sec, respectively, when the fitting procedure is
undertaken in different intervals of $B$ within the range $0<B <0.5 B_{tr}$.
Thus, the accuracy of determination of $a$ and $\tau _{\varphi }$ values is
$20-25$\%. The ratio $ \tau /\tau_{\varphi}$ for this structure is
$0.004-0.009$ for the temperature range $1.5-4.2$ K.

Analogous data treatment for structure II gives  $a=0.6-0.7 $, $\tau _{\varphi
}=0.47\times 10^{-11}$ sec for $T=1.5$ K and the ratio $ \tau
/\tau_{\varphi}=0.03-0.2$ for the temperature range $0.43-4.2$ K.  In the
strict sense the expression (\ref{eq3}) is not valid for this structure
because the scattering potential is smooth and the scattering is anisotropic.
Nevertheless it provides a good agreement with the experimental data (Fig.1
b). It is commonly believed that lower than unity value of $a$ results from
the electron-electron interaction (Maki-Tompson term). However, below it is
shown that such value of  $a$ in structure II is the result of failure of the
diffusion approximation due to poor $\tau/\tau_\varphi$ ratio.

\begin{figure}
 \epsfclipon
 \epsfxsize=\linewidth
 \epsfbox{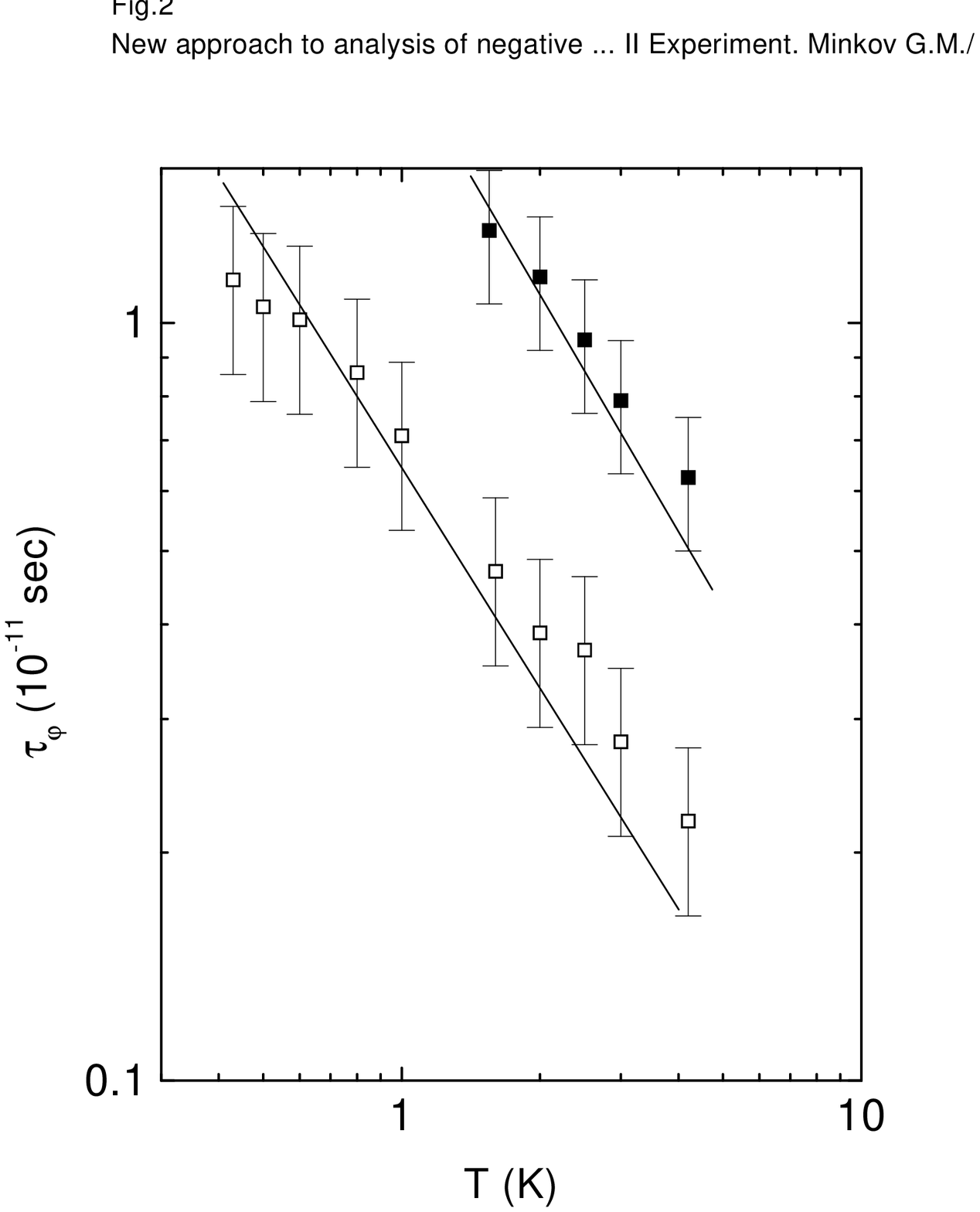}
\caption{Temperature dependencies of $\tau _{\varphi }$ for structure I (solid
squares), and for structure II (open squares). The lines are the dependencies
$\tau _{\varphi } \propto 1/T$.} \label{fig2}
\end{figure}

The temperature dependencies of $\tau _{\varphi }$  are plotted in Fig.\
\ref{fig2} for both structures, and as is seen $\tau _{\varphi }\propto T^p$
with $p\simeq -1$. This means that the inelasticity of electron-electron
interaction is the main mechanism of the phase relaxation.\cite{7}

Now we demonstrate new possibilities of analysis of experimental data provided
by the method described above. It is  obvious that the method is applicable
for low magnetic fields, $B<B_{c}$, where the interference contribution to the
negative magnetoresistance is dominant.  As is seen from Eq.(\ref{eq6}) the
information about the statistics of closed paths can be obtained from the
Fourier transform of $\delta \sigma (B)= \sigma(B)-\sigma(\infty)$. But the
value $\Delta \sigma(B)= \sigma(B)-\sigma(0)$, not $\delta \sigma(B)$, is
experimentally measured. It is easily shown that the Fourier transform of the
experimental  curve $\delta \sigma'(B)= \sigma(0) -\sigma(B_{c}) +
\Delta\sigma(B)$ padded with zeros at $B>B_{c}$ is close to that of $\delta
\sigma(B)$ at $S>\Phi_0/B_{c}$.

\begin{figure}
 \epsfclipon
 \epsfxsize=\linewidth
 \epsfbox{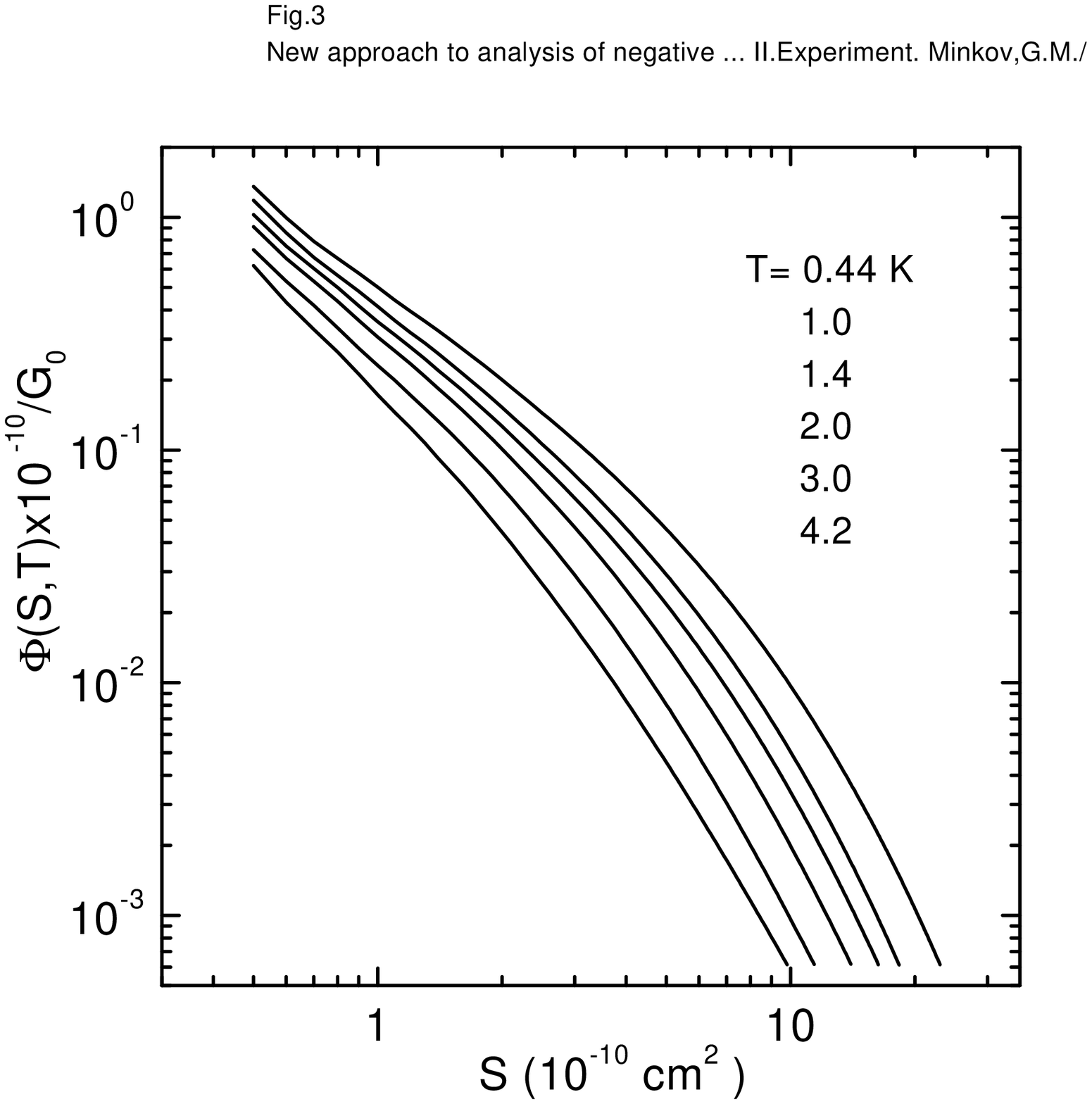}
\caption{The Fourier transforms of $\delta\sigma'(B)$ for
 structure II at different temperatures.}
\label{fig3}
\end{figure}

In Fig. \ref{fig3} the Fourier transforms of $\delta \sigma' (B)$ for
different temperatures are presented. The area range where the Fourier
transform $\Phi(S)$ is shown is bounded. The maximum value of $S$ is
determines by signal-to-noise ratio in the experimental curves $\Delta \sigma
(B)$. The minimum value is determined by the range of magnetic field, $B<
B_{c}$, where the weak localization effect is dominant. We assume that $B_{c}$
is about $0.5$ T for structure I and $0.2$ T for structures II.
\begin{figure}
 \epsfclipon
 \epsfxsize=\linewidth
 \epsfbox{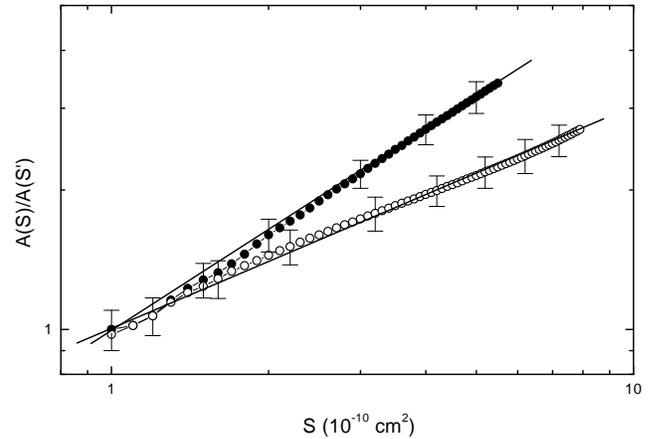}
\caption{Area dependencies of $A$ normalized to $A$ at $S'=10^{-10}$ cm$^2$,
averaged for different temperature pairs. Solid and open circles are the
results for structure I and II, respectively. The upper and lower straight
lines have the slopes $\beta=0.7$ and $\beta=0.52$, respectively.}
\label{fig4}
\end{figure}

As is seen from Eq.\ (\ref{eq7}), the function $\log(A(S))$ must be linear
with respect to $\log(S)$ with the slope $\beta$. It is seen from Fig.\
\ref{fig4} that the corresponding curves are really close to straight lines
for both structures, but the slopes are different: $\beta=0.70\pm 0.05$ for
structure I and $\beta=0.52\pm0.05$ for structure II. Thus, for structure I
with $\tau/\tau_\varphi<0.01$  the value of $\beta$ is close to that obtained
in the improved diffusion approximation $\beta=0.67$, but somewhat larger than
the value $\beta=0.62$ obtained in the numerical simulation.\cite{prev}

In structure II the ratio  $ \tau/\tau _{\varphi } $ is significantly larger
and the diffusion approximation fails. Besides, the fact that the impurities
are arranged in the barriers in this structure leads to smooth scattering
potential and anisotropic scattering. To our knowledge, there are no
theoretical results for this case. However, it is valid to say that the
anisotropy of scattering and smooth scattering potential do not change the
statistics of closed trajectories with the lengths significantly larger than
the mean free length. The numerical calculations beyond the diffusion
approximation for isotropic scattering \cite{prev} show that $\beta=0.55$ at
$\gamma=0.1$. It is close to $\beta$ value for structure I with
$\gamma=0.2-0.03$. Thus we believe that the main reason for small value of
$\beta$ in this structure is failure of diffusion approximation.
\begin{figure}
 \epsfclipon
 \epsfxsize=\linewidth
 \epsfbox{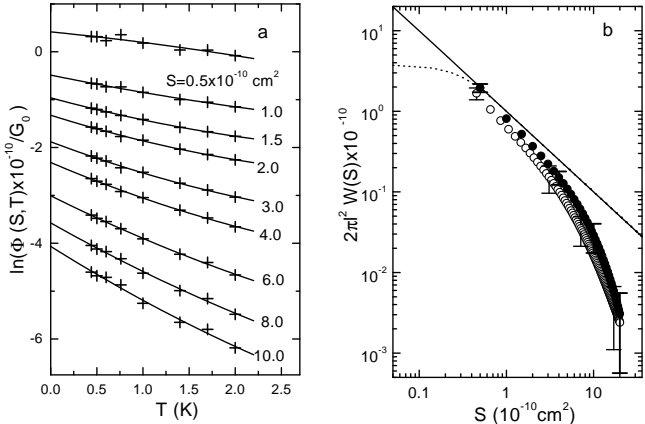}
\caption{(a) The temperature dependencies of $\Phi$ at different $S$ for
structure II. The solid curves show the extrapolation of these dependencies to
$T=0$. (b) The area distribution functions for structure I (solid circles) and
for structure II (open circles). The solid and dotted curves are the area
dependencies of $2\pi l^{2}W(S)$ obtained in the improved diffusion
approximation for structure I and II, respectively. } \label{fig5}
\end{figure}

The method put forward in this paper gives a possibility to determine the area
distribution function $W(S)$. The temperature dependencies of $\Phi (S,T)$ for
several $S$ are plotted in Fig.\ \ref{fig5}a. The value of $\Phi$ for a given
$S$ increases when $T\to 0$ due to increase of  $l_{\varphi }$. Thus the
extrapolation of curves $\Phi(S,T)/G_0$ to $T=0$ (see Eg. (\ref{eq6})) gives
the value of $2\pi l^{2} W$  for the corresponding value of $S$. The results
of such a data treatment for both structures are shown in Fig.\ \ref{fig5}b.

In Fig.\ \ref{fig5}b  the area distribution functions obtained within the
improved diffusion approximation\cite{prev} with parameters corresponding to
the structures investigated are presented too. One can see that at low $S$ the
experimental area dependencies of $2\pi l^{2}W$ are close to the theoretical
ones for both structures. For $S>(4-5)\times 10^{-10}$ cm$^2$ the more rapid
decreasing of the experimental curves is observed and for $S\simeq 10^{-9}$
cm$^2$ the experimental values of $2\pi l^{2}W(S)$ is $3-5$ times lower than
the theoretical values.

There are two reasons for such a discordance: (i) the number of closed
trajectories with large areas in real samples is smaller than the theoretical
one due to, for instance, the long-range potential fluctuations; (ii) the
saturation of the phase breaking length with decreasing temperature from
$T\simeq 1$ K has to lead to underestimating the value of $2\pi l^{2}W(S)$ for
large $S$ in the data processing described above. It should be noted that some
evidence for $l_\varphi$ saturation was obtained only for temperatures
$T<0.15$ K. \cite{102} Our measurements were carried out at significantly
higher temperature, $T>0.4$ K. So, we believe that the rapid decreasing of
$2\pi l^{2}W(S)$ for $S>(4-5)\times 10^{-10}$ cm$^2$ (Fig.\ \ref{fig5}\ b)
results from the shortage of large trajectories, rather than from the
saturation of $l_\varphi$.

Thus, one can see that the area distribution functions of closed paths
coincide practically for both structures, but the area dependencies of the
average length of closed trajectories are distinguished. This distinction
results from different $\tau/ \tau_{\varphi}$ ratio. Just this fact leads to
the lower than unity value of prefactor $a$ in structure II rather than the
electron-electron interaction.

\section{Conclusion}
\label{secconcl} The new method of the analysis of negative magnetoresistance
is used. This method provides a possibility to obtain an information on the
statistics of closed paths. The experimental studies of negative
magnetoresistance show that the area dependence of  average length of closed
paths depends on $\tau/ \tau_{\varphi}$ ratio: $\overline{L(S)}\propto
S^{0.7}$ at $\tau/ \tau_{\varphi }< 10^{-2}$; $\overline{L(S)}\propto S^{0.5}$
at $\tau/ \tau_{\varphi }\simeq 10^{-1}$. This fact leads to the  lower than
unity value of prefactor when one fits the experimental results to Hikami
expression rather than contribution of electron-electron interaction
(Maki-Tompson term). The experimental area distribution functions of closed
paths are close to these obtained in improved diffusion approximation at low
area, but distinct at large one. The shortage of large trajectories by long
range potential fluctuation is a possible reason for such distinction.

This work was supported in part by the RFBR through Grants 97-02-16168,
98-02-17286, the Russian Program {\it Physics of Solid State Nanostructures}
through Grant  97-1091, and the Program {\it University of Russia} through
Grant 420.

\end{multicols}

\begin{references}
\bibitem[*]{cont} Email: Grigori.Minkov@usu.ru

\bibitem{1} S. Hikami, A. Larkin and Y. Nagaoka, Prog. Theor. Phys.
{\bf 63}, 707 (1980)

\bibitem{2} B. Altshuler, D. Khmelnitskii, A. Larkin and P. Lee,
Phys. Rev. B {\bf 22},
5142 (1980)

\bibitem{3} B. Altshuler, A. Aronov, A. Larkin and D. Khmelnitskii,
JETP {\bf 81}, 768
(1981)

\bibitem{102} W.\ Poirier, D.\ Mailly and M.\ Sanquer, cond-mat/9706287;
P.\ Mohanty, E.M.\ Jarivala and A.\ Webb, Phys.Rev.Lett. {\bf 78},
3366 (1997)

\bibitem{prev} G. M. Minkov, A. V. Germanenko,
V. A. Larionova, S. A. Negashev, and I. V. Gornyi, Phys. Rev. B, the previous
paper in this issue

\bibitem{Knap}W.\ Knap, A.\ Zduniak, L.\ H.\ Dmowski,
S.\ Contreras and M.\ I.\ Dyakonov, Phys. Stat. Sol. (b) {\bf
198}, 267 (1996).

\bibitem{Polyanskaya}A. M. Kreshchuk, S. V. Novikov, T. A.
Polyanskaya and I. G. Savel'ev, Fiz. Tekn. Polupr. {\bf 31}, 459
(1997)

\bibitem{7} B. Altshuler, A. Aronov and D. Khmelnitskii,
 J.\ Phys.\ C {\bf 15}, 736 (1982)


\end{references}
\end{document}